\documentclass[nonacm]{acmart}

\AtBeginDocument{%
  \providecommand\BibTeX{{%
    \normalfont B\kern-0.5em{\scshape i\kern-0.25em b}\kern-0.8em\TeX}}}

\usepackage{array}
\usepackage{makecell, afterpage,graphicx}

\definecolor{myblue}{RGB}{0,162,255}
\usepackage{tikz}
\newcommand\encircle[1]{%
  \tikz[baseline=(X.base)] 
    \node (X) [draw, shape=circle, inner sep=0, fill=myblue, text=white] {\strut #1};%
}

\newcommand\revision[1]{{\color{black} #1}}
\newcommand\minorrev[1]{{\color{black} #1}}

\usepackage{comment}
\usepackage{csquotes}
\usepackage{todonotes}
\usepackage{multirow}
\usepackage{dirtytalk}
\usepackage[nolist,nohyperlinks]{acronym}
\begin{acronym}
    \acro{AI}{Artificial Intelligence}
    \acro{HCI}{Human-Computer Interaction}
    \acro{CSCW}{Computer-Supported Cooperative Work}
    \acro{UI}{User Interface}
    \acro{ML}{Machine Learning}
    \acro{SMM}{Shared Mental Model}
    \acro{DS}{Data Science}
    \acro{SMMs}{Shared Mental Models}
\end{acronym}

\begin{document}

\title{How AI Developers Overcome Communication Challenges in a Multidisciplinary Team: A Case Study} 

\author{David Piorkowski}
\affiliation{%
  \institution{IBM Research}
  \country{United States}
}
\email{djp@ibm.com}

\author{Soya Park}
\affiliation{
  \institution{Massachusetts Institute of Technology}
  \country{United States}
}
\email{soya@mit.edu}

\author{April Yi Wang}
\affiliation{
  \institution{University of Michigan}
  \country{United States}
}
\email{aprilww@umich.edu}

\author{Dakuo Wang}
\affiliation{%
  \institution{IBM Research}
  \country{United States}
}
\email{dakuo.wang@ibm.com}

\author{Michael Muller}
\affiliation{%
  \institution{IBM Research}
  \country{United States}
}
\email{michael_muller@us.ibm.com}

\author{Felix Portnoy}
\affiliation{%
  \institution{IBM}
  \country{United States}
}
\email{fportno@us.ibm.com}

\renewcommand{\shortauthors}{Piorkowski et al.}

\begin{abstract}

The development of AI applications is a multidisciplinary effort, involving multiple roles collaborating with the AI developers, an umbrella term we use to include data scientists and other AI-adjacent roles on the same team. During these collaborations, there is a knowledge mismatch between AI developers, who are skilled in data science, and external stakeholders who are typically not. This difference leads to communication gaps, and the onus falls on AI developers to explain data science concepts to their collaborators. In this paper, we report on a study including analyses of both interviews with AI developers and artifacts they produced for communication. Using the analytic lens of shared mental models, we report on the types of communication gaps that AI developers face, how AI developers communicate across disciplinary and organizational boundaries, and how they simultaneously manage issues regarding trust and expectations.
\end{abstract}

\maketitle

\section{Introduction}

As machine learning (ML) continues to transform the world around us, data science teams are becoming a standard fixture at companies and organizations. 
These data scientists provide the expertise necessary to incorporate \revision{ML and artificial intelligence (AI)}  solutions\footnote{In this paper, we use the term \ac{AI} to include both machine learning and artificial intelligence.} into a wide variety of business problems (e.g., healthcare~\cite{doctorbot}, HR~\cite{AllWork:Liao:2018:WNP:3173574.3173577}, and education~\cite{xu161same}).
New technologies and processes for solving problems bring not only new opportunities, but new challenges as well.
Correspondingly, a healthy amount of research has emerged looking at a broad range of topics including the challenges \revision{of} building \ac{AI} software~\cite{amershi2019software, watanabe2019preliminary, wan2019does, arpteg2018software} and how data scientists collaborate~\cite{wang2019data, wang2020callisto, zhang2020data}.

\revision{
Prior work on data scientists' collaborative practices have provided high-level accounts on how data scientists work, and suggested that data science teams face \minorrev{intra-role} communication gaps in code reading, code reuse, and code documentation activities~\cite{kery2018story, rule2018exploration, zhang2020data}.
In this study, we focus on the inter-role communication gap.}
We ran a case study of a data science team at \minorrev{IBM} 
whose role is to provide AI solutions for other product teams in \minorrev{IBM}.
For the purposes of this paper, we term members of this team \textit{AI developers}, since they include members who are AI-adjacent in that they are not data scientists but still require AI knowledge to do their job. 
To help frame our understanding of both the challenges and the solutions, we used \revision{\emph{shared mental models} (SMMs)}  theory~\cite{converse1993shared}. 
At the highest level, \revision{SMMs} provide an abstraction for team members' common understanding of task responsibilities and what the corresponding information needs are. 
This allows them to more readily anticipate each others' needs and work together more efficiently~\cite{mohammed2010metaphor}. 
Effective communication is a core \revision{tenet} of SMMs and the aspect we were interested in using to understand our findings.
In particular, we focus in on the principles proposed by Scheutz et al. for realizing SMMs, which are described in detail later in the paper~\cite{scheutz2017framework}.
\revision{For simplicity in this first paper, we focused only on the AI developer's mental models  \textit{within} a data science team.
Subsequent research should examine the "other side" or "sides" of the communications gap(s), namely the perspectives of other stakeholders in the broader AI product context.}

\revision{
Accordingly, we conducted a semi-structured interview study with four AI expert participants over multiple sessions (in total 10 sessions) to construct a detailed understanding of the communications gaps they faced in the \ac{AI} lifecycle~\cite{automationsurvey} and how they overcame them. Interviewees who participated in two or more sessions additionally shared and discussed communication artifacts 
with us. As a result, we had ten interview sessions in total.}
Given the tight focus and small participant count, we consider this as a formative, yet in-depth, look at the following questions.

\begin{enumerate}
    \item What are the key communication gaps AI developers faced?
    \item \revision{How do AI developers overcome} those gaps and communicating across roles?
\end{enumerate}

In this paper, we detail how AI developers overcome communication gaps using the tools and techniques that are currently available. 
We explain why these approaches are successful by viewing them through the lens of SMMs and bring out the most problematic communication gaps that AI developers face.
In doing so, we provide a glimpse into the difficulties faced by
\revision{people in this role}.
This understanding can then serve future work that aims to address the communication challenges that they face \revision{and the complementary challenges that their colleagues face across the communcications gap, in roles that are outside of the data science team}. 

We found that data scientists 
\revision{are concerned about} communication gaps stemming from mismatched expertise \revision{between the team and their stakeholders}. 
To address the gap, the teams hold informal education sessions and have continuous sync-up sessions to establish trust. 
Finally, we discuss best practices in AI development according to SMMs.
\revision{\section{Related Work}}
Our research contributes to the growing discussion of collaborative data science in \ac{HCI} and \ac{CSCW}
(e.g., \cite{aragon2016developing, qazi2019interactive, kogan2020mapping, muller2019human}).
We summarize related work into four parts: (1) software engineering for \ac{AI} systems, (2) emerging roles and collaboration practices of an \ac{AI} team, (3) communication challenges in \ac{AI} teams and software engineering teams, and (4) SMM theory. \\

\subsection{Software Engineering for \ac{AI} Systems}
The advances of \ac{AI} techniques open the opportunities for \minorrev{companies} and developers to build software products that handle tasks intelligently, improve accuracy and efficiency, and provide personalized experience.
\ac{AI} has been widely applied to many domains such as helping diagnose patients~\cite{ doctorbot,aidoctor}, building chatbots to improve customer service~\cite{chatbot_service_agent_Xu:2017:NCC:3025453.3025496,AllWork:Liao:2018:WNP:3173574.3173577}, and enhancing the education experience with intelligent tutoring systems~\cite{xu161same}.

Building \ac{AI} products involves not only the process of extracting insights from data (data-centric), but also constructing models from insights and building software products (model-centric or product-centric).
The process of extracting insights from data has been well-studied and categorized into three high-level stages~\cite{wang2019hai}: data preparation, model building, and model deployment.
While some studies in \ac{HCI} and \ac{CSCW} investigated the workflow from a data-centric perspective~\cite{muller2019data, kougka2018many, passi2018trust}, others discussed the process of integrating models and building \ac{AI} products in detail~\cite{amershi2019software, watanabe2019preliminary, wan2019does, arpteg2018software}, 
Amershi et al.~\cite{amershi2019software} summarized nine stages of machine learning workflow as data-oriented (e.g., collection, cleaning, and labeling) and model-oriented (e.g., model requirements, feature engineering, training, evaluation, deployment, and monitoring).
This workflow is iterative and contains many feedback loops.
In practice, \ac{AI} developers often struggled to establish a repeatable process, as Hill et al. found in their interview study with AI developers from various expertise levels~\cite{hill2016trials}. In this paper, we use the AI development lifecyle as proposed by Amershi et al.'s work ~\cite{amershi2019software}.

Prior research investigated the technical challenges of software engineering for \ac{AI} systems, such as difficulties in problem formulation and specifying desired outcome, lack of critical analysis of training data, and lack of evaluation of models with business-centric measures~\cite{washizaki2019studying, lwakatare2019taxonomy}.
In addition, building \ac{AI} products requires the development team to consider many facets, such as interactions, performance, cross-platform experience, and social good~\cite{sukis_2019, hill2016trials}.
Thus, the \ac{AI} development team must embrace collaboration between different roles and leverage expertise from each other---despite the relatively low usage of documentation for this purpose \cite{pine2015politics, rule2018exploration, zhang2020data}.
Our work aims to reveal the hows and whys of the \ac{AI} development teams communicating across roles and stages.

\subsection{Emerging Roles and Collaboration Practices of \ac{AI} Development Teams}
Prior research has shed light on emerging roles and collaboration practices in a data-centric workflow.
We use the phrase ``data scientist'' referring to people who have the technical skills to find trends and manage data in a variety of domains, such as decision sciences and business intelligence, product and marketing analytics, fraud and risk analytics, data services and operations, and data engineering and infrastructure~\cite{patil2011building}.
Building \ac{AI} systems falls into the criteria of data engineering and infrastructure, where development and implementation skills are critical for data scientists.
With the growing impact of data science, many jobs across all industries continue to be changed by data scientists~\cite{van2014data}.
For example, many software developers are now learning machine learning through informal resources (e.g., interactive machine learning tools \cite{yang2019unremarkable}), hoping to adopt machine learning into their own practices \cite{cai2019software}.

Effective collaboration can help data science teams to leverage expertise from each other and to improve quality of work~\cite{wang2019data}.
Studies have investigated collaboration models among different team settings, such as professional data scientists~\cite{wang2019data, wang2020callisto}, civic data scientists~\cite{hou2017hacking}, domain experts~\cite{mao2019data}, and software-oriented data analysts~\cite{kim2016emerging}.
Zhang et al. used a survey to study the general collaboration practices of data science teams among both technical team members and non-technical team members\revision{~\cite{zhang2020data}}.
Building on top of Zhang et al.'s work, our work takes a deeper focus and specifically examines kinds of communication gaps and collaboration practices in \ac{AI} development teams. \minorrev{Through} in-depth interviews, we discovered that mismatched expertise between data scientists and business experts hampers their collaboration\minorrev{,} and \minorrev{we learned how data scientists} address the issue.

\revision{
\subsection{Communication Challenges in \ac{AI} and Software Engineering Teams}

\minorrev{Prior work has also investigated} the communication challenges that happen in \ac{AI} and software engineering teams. For example, Storey et al. found that engineer \minorrev{users} on GitHub have a wide variety of ``socially enabled communication channels'' and social media to exchange information along with their coding project, but they still faced communication challenges such as miscommunication, language barriers, fear of intimidation and poor attitudes ~\cite{storey2016social}. 

In the data science domain, the communication challenge is known to be broad and fraught with difficulties. Mao et al.~\cite{mao2019data} reported that in scientific discovery projects, data scientists and bio-medical scientists often have different motivations. Such mismatch of their the motivations complicated their communication behavior in that bio-medical scientists users kept asking data scientists new questions, even though the data scientists had not yet found an answer for their previous question. Hou et al. reported \minorrev{in} their interview study 
\minorrev{that} subject matter experts and data scientists 
''speak a different language''~\cite{hou2017hacking}. Sometimes the data scientists do not know how to translate a subject matter expert's business problem into a well-defined data science problem. Thus, it may need a third party to be the bridge or the ``broker role''~\cite{williams1993translation} to do the translation for the communication.

To remediate these \minorrev{communication} challenges in an AI or software development team, prior works in \ac{HCI} have proposed novel collaboration systems to aid the teamwork~\cite{bhardwaj2014datahub, wang2020callisto, storey2006shared, treude2009empirical}.
For example, Wang et al. built a system that allows AI developers to chat by the side of their code in a Jupyter Notebook environment, and the system can use those informal communication messages as contextual links for the code~\cite{wang2020callisto}. Storey et al. investigated shared waypoints and social tagging in collaborative software development~\cite{storey2006shared}. But most of systems were built for supporting communications between the technical AI developers, not considering the inter-role communications between an AI developer and a domain expert.

Thus, we conducted this interview study to investigate AI development teams' communication practices, with a focus on the communication challenges that happened between the AI developers and the domain expert collaborator. By reflecting on the tools and techniques that \ac{AI} developers use to communicate with other stakeholders, we hope to open opportunities for future tool designers to build better collaboration tools.

}

\begin{table}
\caption{A summary of the principles of how humans realize shared mental models \revision{\cite{scheutz2017framework}. This framework is helpful for describing collaborations activities in the domain of AI.} 
    }
    \begin{tabular}{p{2cm}p{11cm}}
    \toprule
    \textbf{Principle}  & \textbf{Definition} \\ 
    \midrule
    Consistency         & Ability to maintain stability by resolving conflicts due to differing perceptions, differing knowledge states, asynchronous information, and missing updates. \\ 
    Reactivity          & Ability to effectively address unanticipated events or state changes by informing team members of the changes and adapting goals and plans to account for the new situations. \\ 
    Proactivity         & Ability to anticipate problems, bottlenecks and failures and take proactive actions, such as asking for clarification or offering assistance.                                 \\
    Coordination        & Ability to work together via overall cooperative attitudes, such as establishing joint goals and plans, transparent task assignments, and truthful information sharing        \\ 
    Knowledge Stability & Ability to understand the staleness of information over time and adjust sampling rates according to their confidence in the information’s validity.                     \\
    \bottomrule
    \end{tabular}
    
    \label{tbl:smm-principles}
\end{table}

\subsection{Shared Mental Model}
Effective team collaboration, particularly multidisciplinary collaboration, requires team members to hold SMMs on task requirements, procedures, and role responsibilities \cite{converse1993shared}.
Studies revealed that in multidisciplinary collaboration, professionals are unwilling to accept overlapping duties over roles \cite{jones2006multidisciplinary} and rely on collaborative work tools to establish common ground \cite{adamczyk2007supporting}.

Inspired by the general literature on common ground \cite{clark2006context}, our work uses SMM theories as a lens to discuss the communication and collaboration practices between AI developers and other stakeholders. 
In this approach, an SMM provides one means to establish common ground \cite{van2011team}.
The SMM literature distinguished between two models of shared understandings across team members \cite{lee2012toward, scheutz2017framework, resick2010mental}.
The \textit{task model} consists of equipment (e.g., equipment functioning, operating procedures) and task (e.g., task procedures, task strategies), while the \textit{team model} consists of team interaction (e.g., roles, communication channels) and team (e.g., knowledge, skills).
Yu et al. \cite{yu2014understanding} summarized four stages for developing an SMM within agile software development teams from prior literature: knowing meta-knowledge of the team \cite{mccomb2007mental, larson1994discussion}, learning and building the team’s transactive memory system \cite{austin2003transactive}, understanding and reaching consensus \cite{van2011team}, and executing team goals.
They identified several activities in agile software development practices that can improve collaboration using SMM theory (e.g., planning, reflexivity, leader briefing -- which are integrated in our coding schema in Table ~\ref{tbl:code-list}).
Scheutz et al. extended SMMs into the domain of AI, using SMMs as the theoretical foundation behind frameworks describing collaborations activities of humans and agents~\cite{scheutz2017framework}. Grounded in prior work on SMMs, they proposed five principles of how humans realize SMMs, shown in Table \ref{tbl:smm-principles}.
We leverage these principles as the lens through which we explain and understand of how \ac{AI} developers work.
Our work aims to uncover the unique practices and challenges that \ac{AI} developers have when building SMMs with other stakeholders.

\section{Methodology}

\minorrev{To identify the communication gaps that AI developers faced, we interviewed AI developers and qualitatively analyzed interview transcripts them using a SMM lens. What follow are the details of both the participants and the study design.}

\subsection{Participants}
\label{subsect:participants}

\begin{table}[b]
 \caption{Interview participants, labeled based on their role in their team. Expertise and the number of project collaborators are provided by the participants. * We reviewed artifacts with Data3, but due to confidentiality, we are unable to share information from these interviews in this work. Data3's project overview interview data is included.}
  \begin{tabular}{p{2.2cm}cp{4.5cm}p{2.4cm}c}
  \toprule
  \thead{Role\\ $[Label]$}&\thead{Years of\\ experience} 
  &\thead{Expertise}&\thead{Project\\ collaborators} & \thead{ \# Artifacts \\shared}\\
        \midrule
        Data scientist [\emph{Data1}] & 4 & Machine learning, unsupervised learning, natural language processing, and statistics  & 8 internal, 10 external & 2\\
        \midrule
        Data scientist [\emph{Data2}] & 4 & Building data science models in a business setting to help business transformation & 10 internal, about 10 external & 1\\
        \midrule
        Data scientist [\emph{Data3}] & 2 & Machine learning, deep learning, statistical analysis & 7 internal, 2-5 external & $0^{*}$ \\
        \midrule
        Strategy consultant [\emph{Strat1}] & 2 & Business strategy and project implementation & 3 internal, 8 external & 5\\
      \bottomrule
    \end{tabular}
   
    \label{tbl:participants}
\end{table}

\revision{Our study involved real projects in \minorrev{IBM}.
We wanted to understand difficulties and pain points of production-targeted data science work-practices.
Participants described issues and experiences that could only be communicated with agreed protections of privacy and confidentiality. Therefore, our sample was restricted to 
\minorrev{within the}
company.}

\minorrev{Using our access as employees of the same company, we}
recruited participants 
\revision{through}
an established working relationship with the data science team represented in this study. 
We will refer to this team as the \textit{AI team}.
We sent out emails to 
\minorrev{IBM's}
\minorrev{mailing lists} detailing the selection criteria. 
The criteria were: 
\begin{itemize}
    \item the participant had to be working on a project involving \ac{ML} or \ac{AI}
    \item the project was nearing completion or complete
    \item \ac{ML} or \ac{AI} was an important part of their role on the project
\end{itemize}
Our recruitment resulted in four participants from the same department in the company, but from different teams. 
Table \ref{tbl:participants} shows participants' roles, experience and expertise. 
The brackets after each role refer to the labels we use to identify each participant's quotes.

We established a working relationship with \revision{these teams} specifically because of a couple \revision{of} key aspects of how they work. First, they work on a wide variety of problems with a variety of teams, providing them with a perspective that is wider than a data science team working on a single domain. 
Furthermore, they do not maintain any of the \ac{AI} applications they build. 
Instead, they hand off all the projects they create for other teams to maintain.
Therefore, the \ac{AI} team should be incentivized to \revision{provide} enough information to others for maintainability's sake.
Thus, we hoped that this aspect would bring to light useful collaboration practices. 

\subsection{Interview Protocol}
To answer our research questions we ran a total of 10 hours of semi-structured interview sessions over four participants. We split interviews into two different phases lasting from 45 minutes to one hour:
\begin{itemize}
    \item The \textit{project-overview phase} consisted of one interview to describe the \ac{AI} project and roles involved. It also served to scope future interviews. \revision{We conducted four project-overview sessions over four participants.}
    \item The \textit{artifact-in-depth phase} consisted of interviews to discuss one or more artifacts brought by the participant to demonstrate how they communicate across roles. \revision{We conducted six artifact-in-depth sessions over four participants, with interviewees sharing one to three artifacts per interview.}
\end{itemize}

In the project-overview interviews, participants recalled an \ac{ML} or \ac{AI} project they had recently finished or were about to finish. 
In these interviews, we asked participants about details of the project, the composition of the team, and how ML or \ac{AI} fit into the project. 
At the conclusion of the project overview interview, the researcher and participant jointly built a visualization in Mural\footnote{https://mural.co} which aimed to map the various roles involved on the project to the phases of the \ac{AI} development lifecycle. 
Figure \ref{fig:workflow-role} shows a consolidated view of these joint visualization activities.
In this task, participants were shown a proposed starting \ac{AI} development lifecyle and were invited to modify it to make it more closely match the project they described. 
Once they finished modifying the lifecycle, participants were asked to place the roles that were mentioned during the interview into any of the development phases that were represented in the Mural. 
Roles included members from the participant's team, and also other teams who were involved in the project. 
In our case, all four participants were working with at least one \say{client} team, whom we term the \textit{Stakeholder team}, that would eventually take on ownership and maintenance of the final product.

In the artifact-in-depth interviews, participants shared artifacts that had previously or currently served as a mechanism for facilitating communication across roles.
We asked participants to share artifacts used in \minorrev{either of} the \minorrev{``M}odel evaluation\minorrev{'' or ``Model} deployment \minorrev{and integration''} \ac{AI} lifecycle stage\minorrev{s} if possible.
\minorrev{We chose these two stages because, according} 
to the project-overview interviews, \minorrev{these two stages}
tended to involve lots of different roles across multiple teams, and thus, 
\minorrev{were}
likely to reveal communication challenges.
\minorrev{Additionally, AI developers indicated in the project-overview interviews that these stages required them to produce explanatory artifacts.}
Participants shared from one to five artifacts with researchers, which were discussed in detail in the artifact in-depth interview sessions.
Artifacts
consisted of digital objects such as slide presentations, documentation, software repositories, and README files. For each one of these artifacts, we ran an artifact-in-depth interview that uncovered how the artifact was created, how it was used, how successful it was, and its limitations.
One participant's artifacts and corresponding interviews were removed from the study due to confidentiality concerns, but data from the project-overview interview remains.

Interviews were conducted remotely using WebEx\footnote{https://www.webex.com}, were recorded and transcribed for qualitative analysis.

\subsection{\ac{AI} Team Working Practices}

\begin{figure}[h]
      \centering
      \includegraphics[width=\linewidth]{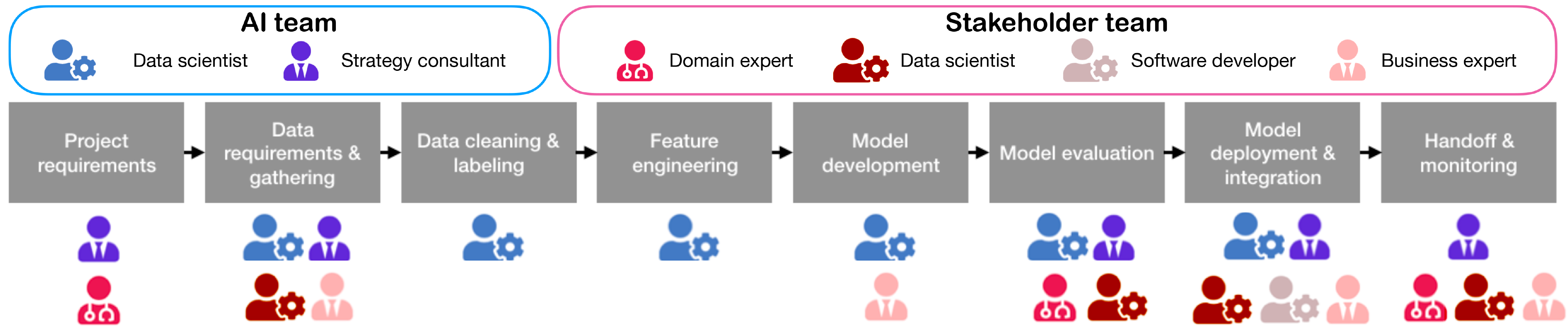}
      \caption{Aggregated view of our interviewees' different roles involved in each stage of \ac{AI} workflow. \revision{Each team's composition slightly varies, and can be found in the Appendix.}  An \ac{AI} team consults and implements \ac{AI} solutions. It normally consists of technical and business representatives. A Stakeholder team requests help from a \ac{AI} implementation team. They are knowledgeable in their domain, however lack knowledge in \ac{ML} or \ac{AI}. \revision{We presented the stages linearly, but in reality the AI development process is highly iterative.}}
      \label{fig:workflow-role}
    \end{figure}

\footnotesize
\begin{table*}[] 
\caption{The coding book (9 categories and 27 concepts) generated from iterative axial coding. }
\begin{tabular}{p{2.8cm}p{3.5cm}p{6cm}}
\hline
\textbf{Category} & \textbf{Concept} & \textbf{Definition (Utterances about...)}\\ \hline
\multirow{5}{2.5cm}{Data Scientists Provided Information} & Objective & ... project objectives \\  
 & Algorithm selection & ... deciding between algorithms and final algorithm selection \\  
 & Model performance and results & ... evaluating model quality, model performance or model results. \\  
 & Model improvements & ... model improvements when compared to a previous version or an existing baseline \\ 
 & Model validation & ... validating model results \\ \hline
Stakeholders Requested Information & Stakeholder requested information & ... information needs requested by the Stakeholder team\\ \hline
\multirow{5}{2.5cm}{Data Scientists' Rationale for Providing Certain Type of Information} & Missing info & ... reasons to address missing information \\  
 & Data science education & ... reasons to educate others about data science concepts \\ 
 & Contextualization & ... reasons to contextualize the results in the problem domain\\ 
 & Legal and regulatory & ... reasons regarding legal or regulatory compliance \\ \hline
\multirow{5}{2.5cm}{Information Conveyance Selection} & Examples & ... using a data instance as an example \\  
 & Visualizations and plots & ... building visualizations and plots \\ 
 & Listing numbers and metrics & ... providing numerical data such as summary statistics or metrics\\ 
 & Explanatory story & ... creating an analogous story to provide explanation\\
 & Definitions & ... defining terms, typically in a less-technical way \\ \hline
Data Science Project Workflow & Data Science Project Workflow & ... steps, tasks, timeline, and other details about the workflow \\ \hline
\multirow{2}{2.5cm}{Data Science Collaboration Roles} & Stakeholder team roles & ... roles and responsibilities of members in the Stakeholder team \\ 
 & \ac{ML} team roles & ... roles and responsibilities of members in the \ac{ML} team\\ \hline
\multirow{2}{2.5cm}{Trust Between \ac{ML} Teams and Stakeholder Teams}& Trust development & ... how trust was built between the two teams\\ 
 & Distrust \newline & ... instances of distrust between the two teams \\ \hline
\multirow{4}{2.5cm}{Communication} & Communication challenges & ... communication difficulties faced by the Stakeholder team \\ 
 & Communication artifacts & ... digital creations to aid communication between teams\\ 
 & Communication frequency & ... how often communication happened between roles\\ 
 & Efforts in preparation for communication & ... timelines or work done to prepare communication artifacts\\ \hline
\multirow{2}{2.5cm}{Technology}& Communication technology & ... technologies used to prepare communication artifacts\\ 
 & Data science technology & ... technology used for data science\\ \hline
\end{tabular}

 \label{table:codebook}
\end{table*}
\normalsize

Participants reported that in general, two teams are involved in the \ac{AI} project: an \ac{AI} team and a Stakeholder team. The \ac{AI} team is responsible for gathering the Stakeholder team's needs, formulating the problem into the data science domain, and building an \ac{AI} model to solve the problem. The Stakeholder team often has deep expertise and domain knowledge in a particular field, and is responsible for helping the \ac{AI} team to understand the real-world problem, providing guidance, and verifying and consuming the final model result.

The \ac{AI} team has two main roles, the data scientist and the strategic consultant. 
The data scientists build and evaluate models. 
The strategic consultants shepherd the project and keep up the communication between the two teams. 
On the Stakeholder team are the domain experts, data scientists, software developers and business experts. 
Domain experts are the subject matter experts in the problem domain that the \ac{AI} team is trying to help solve. 
Although data scientists exist on the Stakeholder teams, their role is the project is to maintain \revision{the} solution that is built by the \ac{AI} team. 
Stakeholder team data scientists tend to be more specialized to their specific business domain then the data scientists on the \ac{AI} team. Software developers are responsible for helping package the \ac{AI} solution into an application or service. 
Finally, the business expert is a management role that oversees an entire line of business that would be represented by multiple domain experts.

An aggregated view of how the roles are sorted and related to the \ac{AI} development lifecycle is presented in Figure \ref{fig:workflow-role}. This figure was generated from the Murals that participants created as part of the project-overview interview. \revision{{This figure shows how the roles drop in and out of different phases of the lifecycle as the project matures.}} As a result, much of the work happens outside the view of others, hinting as to why maintaining a shared knowledge state across all the different roles is challenging.

\subsection{Qualitative Coding}
To help map the current practices for achieving SMMs to the communication gaps that \ac{AI} developers faced, we needed to identify the relevant qualitative data in the transcripts.
To do so, we conducted \revision{multiple interview sessions with our participants.
As a result, we had 10 sessions.
We followed a consensus-coding protocol, as recently reviewed by McDonald et al. \cite{mcdonald2019reliability}. 
We} conducted axial coding of the 10 interview transcripts with an open-coding protocol.
Two of the researchers started with one interview and each developed a code book independently to generate concepts and organize these concepts into categories.
The concepts were chosen to be broadly about topics related to the types of information exchanged, how information was exchanged, and difficulties around information exchange.
The two coders then sat together and discussed the disparity of the code book and their understanding and definition.
Then, they split and independently coded transcripts again based on the agreed upon concept and category definitions.
This process was iterated upon for 4 times, until they come to an agreement about the concepts, categories, and general themes.
Once agreement was reached, they divided and conquered the other nine interview transcripts.
The final code book has nine categories and 27 concepts and is summarized in Table~\ref{table:codebook}.

\section{Results}

Participants faced a varied assortment of challenges related to communication gaps.
The results of our qualitative analysis are presented in two parts. 
First, in Section \ref{sec:challenges}, we bring to focus the three communication gap themes that AI developers faced.
We also elaborate on the role that \minorrev{education} plays in the communication gaps we uncovered.
Second, in Section \ref{sec:overcome}, we dive into the AI developers' world and highlight specific examples of challenges that they faced. 
Those examples are organized using the principles for realizing SMMs presented in Table \ref{tbl:smm-principles}. 
These vignettes highlight the communication gap and the techniques and tools participants chose to address them. 
We also provide real-life examples from slide decks that demonstrate the techniques that AI developers used to get their points across.

\subsection{\minorrev{RQ1:} Kinds of Communication Gaps}
\label{sec:challenges}

Participants' reflections on communication challenges 
centered on three major themes: knowledge gaps across roles, establishing trust, and setting expectations.

\subsubsection{Knowledge Gaps between Roles due to Mismatched Expertise} 
\label{subsubsect:expertise}
Team members came from different technical backgrounds and expertise. This configuration makes the collaboration challenging. ML and AI involve several domain specific concepts such as model algorithms, evaluation metrics and advanced statistical measures 
that make it difficult for software engineers to learn, let alone business stakeholders~\cite{cai2019software}. AI teams struggle to teach these concepts  to stakeholders without getting into too much detail.
The lack of AI knowledge caused stakeholders to misinterpret model performance: \textit{``For defects, .. a lot of times it’s because people don’t understand how the model works''} [Data2].

Teams tend to go through an \emph{education phase} at the beginning of the project to address this issue. The ultimate goal of such education is to bridge the knowledge gap and to lay the groundwork for upcoming collaboration. Echoing findings from previous work, one of the common practices is to hold AI education sessions for software engineers to overcome the knowledge barrier~\cite{amershi2019software}.
AI teams' education efforts on the fundamentals of machine learning extended beyond software engineers, to other non-AI savvy roles such as business users. For example, one kind of education specifically focused on mapping machine learning terminology to business terminology. One data scientist said: \textit{``Some of the languages came from them [business stakeholders]. Like, they had an idea of the near match, which is from their perspective''} [Data1].
However, such education sessions are done in an informal manner and there is no effective way to teach the team, \revision{resulting in lots of} trial and error. [Strat1] said: 
\revision{\textit{``If one way [of explaining] doesn't work, .. where \revision{there's a (need for)} common understanding or common communication. We... really try to keep building examples and education materials along those lines.''}}

This education session is \revision{\emph{bi-directional}}. Domain experts also educate AI teams by explaining domain knowledge and the desired behavior of the model. Domain knowledge is hard to capture in one session. \minorrev{Also topics change as patterns and findings emerge from the data as the teams} progress in the project. Thus, AI teams repeatedly ask questions and confirm to domain experts throughout the entire collaboration.

\subsubsection{Establishing Trust Across Disciplines}
\label{subsubsec:establishingtrust}

Given the nature of a multi-disciplinary team without a shared work history, it is difficult to establish trust between team members~\cite{lacey20029, jones2006multidisciplinary, adamczyk2007supporting}.
\minorrev{Due to these fundamental differences,
when business stakeholders and data scientists work together,} stakeholders do not understand the hurdles and contribution of data scientists and vice versa. According to Strat1, \textit{``We also even schedule additional time beyond that specifically on particular topics that we find difficult just because we know this is a new area and it's kinda key''}.
Without building enough trust upfront, conflicts are inevitable. The same participant said:
\textit{``There's a lot of trust and understanding upfront and when we kinda give these high-level structures and architectures of what we can deliver''} [Strat1].

Trust building took on many forms, not only from repeated contact with regular meetings but also in the varied efforts that the project team undertook to explain particular model concepts or metrics. Additional effort was spent in addressing differences between how AI is developed versus more traditional software. An important example of this was when the AI implementation team explained to the stakeholders that the model's performance does not necessarily improve as they make incremental improvements on a model:
\revision{\textit{``[When] we didn't have any particular performance increase, we had to give an explanation  
(and to let) them know that it's kind of a trial and error... 
That was a real kind of struggle to let them know, we weren't actually doing anything wrong and nothing was broken''
}  [Strat1]. 
}

\subsubsection{Setting and Managing Stakeholder Expectations}
\label{subsubsect:expectations}

Setting expectations played a key role to clients who were unfamiliar with the uncertainties inherent to machine learning:
\textit{``They [stakeholders] had made it very clear previously, that they're not comfortable moving forward with what they deem as a `[closed] box'. So we need to be as transparent as possible in the modeling''} [Strat1]

For instance, at the early stage of collaboration, AI teams demonstrate a sample model to stakeholders and show its prediction on certain data points. The teams have to persuade stakeholders that AI will outperform current techniques being used in their service: 
\begin{displayquote}
\textit{``This team is very used to rigid logic rules or logic trees in terms of making decisions, ...
but a big miscommunication... difficulty [was] they would see one particular data point or factor and say well, this always should give a positive match. 
And we have to really explain that''} [Strat1]\end{displayquote}

Additional challenges of such persuasion stem from the fact that each stakeholder is looking for different performance metrics, hence there is no general way of communication but it makes the AI experts work on a case-by-case basis: \textit{``A lot of times depending on their subjectivity. I think that's the hard part to really assess the benefit --how this model has improved''} [Data2].

\subsubsection{\minorrev{Communication Gaps from a Shared Mental Models Lens}}

\minorrev{SMM theory's principles provide a framework for interpreting how AI developers addressed the communication gaps identified above. By framing the issues from the lens of theory, we begin to uncover potential reasons \emph{why} certain strategies are successful and thus, can better support AI developers in the communication challenges that they face. We summarize the findings above from this perspective as follows.}

\revision{ 
From a \emph{consistency} point of view, the desired outcome from setting expectations and education is to get both the AI and Stakeholder teams on the same page working towards the same purpose. From a \emph{proactivity} perspective, the AI team's past experiences and expectations lead the team to address potential misunderstandings early and also, help build and maintain trust. Finally, from a \emph{reactivity} perspective, part of managing expectations is reconciling misunderstandings when they 
occur. In such situations, the AI team has to be quick on their feet as further explained in Section \ref{sec:overcome}. }

\subsection{RQ2: How AI Developers Cross Communication Gaps}
\label{sec:overcome}

Prior research has shown the kinds of problems faced in developing AI software \cite{hill2016trials, amershi2019software} and the tools used in this space \cite{zhang2020data}, but to our knowledge, there is little known beyond survey results \cite{zhang2020data} about the contextualized specifics of how AI developers communicate with other roles or the motivations behind that communication. 
In this section, we shed light on these questions by providing specific examples of participants' responses to information needs.
We loosely frame the discussion using the principles of how people realize SMMs from Table \ref{tbl:smm-principles} as they neatly capture participants' motivations from the interviews. 

\revision{The coordination principle is omitted due to a lack of data from participants about it.
The coordination principle emphasizes joint goals and plans and information sharing (Table \ref{tbl:smm-principles}).
However, part of the answer to RQ1, above, was that the two teams' mismatched expertise (section \ref{subsubsect:expertise}), resulting in different perspectives and languages about goals and plans, especially in terms of expectations (section \ref{subsubsect:expectations}). Those differences may have reduced the effectiveness of their information-sharing.
Thus, the absence of discussions related to the coordination principle is consistent with what we learned about communication gaps in relation to RQ1.
}

\subsubsection{Consistency Gaps: How Knowledge is Communicated}

At its core, the consistency principle is about maintaining a shared knowledge state.
The reasons for why SMMs fall into conflict vary, but in this case study, we observed effects from the asynchronous nature of the work and the imbalance of data science skill across roles.
Recall that the AI team and Stakeholder teams  mingle, separate and mingle again in the different phases of the lifecycle and that there are entire phases of development where data scientists' skills are put to use building models in (mostly) isolation. 
Consequently, there are abundant opportunities for mental models to differ\revision{. Therefore,} 
in this section, we describe the different ways that AI developers shared information with others, via specific examples of the problems they faced and how they overcame them.

One recurring way that participants overcame the asynchronous nature of their work and maintained consistency was by answering questions from other roles.
Many of the questions fielded by the participants are best described as quick explanations such as explaining what certain lines of code did,
\textit{``Most of the questions were in reference to the actual code like, `in this (Python) file'. Like, `I don't know what line two hundred is doing' ''} [Data1]
or specific requests for code documentation,
\textit{``I mean after several iterations of just communication back and forth I understood that they needed to know kind of more code documentation for the error codes in particular''} [Strat1].
Participants described these types of requests as straightforward to answer as they could be addressed after just a quick glance at the code in question.
Questions such as these were answered
during meetings using screen-sharing technologies, via message services such as Slack, or email. 

As questions started to repeat, participants pivoted to shared information spaces such as cloud content management tools (e.g., Box, Dropbox) or GitHub repositories' README files, Data1 recalled such an instance:
\begin{displayquote}
\textit{``It started out more with just them asking questions and, you know, I would have meetings to respond to the questions, things like that. But after maybe two or three of those meetings, I started to realize that we're kind of getting some questions repeated... so I thought it would just be easier to get all the one document, then we can share that.''}
\end{displayquote}
The consolidation of information into a shared resources can be viewed as a way to reduce the cost of interruption by making the questions publicly available to others (e.g., \cite{muller2009information}).

AI developers' data science expertise often left them as the only ones in the room who could explain what a model was doing and how to interpret it. Knowing this, AI developers took the task of educating others seriously, often spending lots of time preparing materials.
Strat1 recalled one example where he spent time over ``2 to 3 days'' preparing three PowerPoint slides to explain the concepts of precision and recall to the Stakeholder team.
Although the team understood the concepts, they were unable to sufficiently map those concepts to the business problem being solved.
The interviewee lamented, \textit{``They (Stakeholder team) understood recall and precision better, but that still didn't really resonate with them in terms of business impact and more.''}

The situation above is indicative of the overarching educational challenge that all the AI developers talked about: how to map a problem from the business domain to the AI domain.
\revision{This problem is particularly thorny since each group seems to express the problem in their own language (again impacting the substrate on which the coordination principle depends), and this} 
complicates communication regarding how a business problem is translated to a model's algorithm and outputs.
\begin{displayquote}
\textit{``I guess for this audience, they had some concept of how--- what this means conceptually, right? Some of the language came from them. They had an idea of the near match, which is from  their perspective... This was a way to kind of explicitly say how we're identifying it (the concept of a near match) in the code...I don't think they understood how we were combining all these different elements into this measure of closeness''} [Data1]
\end{displayquote}
Data1's presentation attempted to address this gap by mapping the content of his presentation to stakeholder roles' perceptions. He did so by leveraging a specific use case that stakeholders were very interested in solving.
\begin{displayquote}
\textit{``I think these examples were chosen intentionally because for (the Stakeholder team) this (use case) is a really big one for them... So it is something that was on the forefront for them because we thought this was a relevant example for this audience because that was something they were focused on''} [Data1]
\end{displayquote}
We observed that AI developers have a tremendous influence on decisions made by the Stakeholder team due to this knowledge imbalance.
Successes and failures could be traced back to how effectively an AI developer was able to educate others on how a model worked.

\begin{displayquote}
\textit{``I guess the major outcome, which I think we achieved, just try to get management up to speed on how we had improved this algorithm and hopefully also get them excited about it... Our ultimate (goal) is to actually implement this in their business so they can start to leverage it and hopefully make some improvements''} [Data1]
\end{displayquote}

\subsubsection{Addressing Gaps: The Role of Reactivity and Proactivity}

Uncertainty is a given when building an intelligent system, so it becomes an important aspect of maintaining an SMM.
Dealing with that uncertainty unfolds in two ways captured in the reactivity and proactivity principles. Uncertainty leads to changes which require teams to respond. Reactivity is about responding to \textit{unanticipated} problems whereas proactivity is about responding to \textit{anticipated} problems.

Unexpected problems that participants handled reactively included explaining model performance and understanding end users. One of the \minorrev{well established difficulties} of building AI systems is that it is difficult to predict a model's performance without first building the model and evaluating it. 
Techniques for estimating how likely a model is going to meet stakeholder expectations often boils down to AI developers' experience. 
This lack of predictability is often at odds with stakeholders' preexisting experiences with working with traditional software teams.
In \revision{S}ection \ref{subsubsec:establishingtrust}, we described Strat1's difficulties in explaining why models were ``underperforming'' compared to their best shared predictions and they had to explain the discrepancy to their business counterparts through explaining how machine learning evaluations work.

In such instances, participants exhibited patterns of reactive communication, where they would decipher possible causes for under-performance or prepare some alternative approaches to try in the next iteration.
This allowed the team to focus on other issues in the project, as a plan was already in place for the underperforming model.

In another example, Data2 talked about how users were unwilling to use the system that was created because they were unwilling to trust the prediction it made.
They summarized the problem:
\begin{displayquote}
\textit{``The biggest challenge for us, I believe, is we need to communicate with business stakeholders and with all the users, to persuade them, educate them, or prove to them that the [prediction] does make sense''} [Data2]
\end{displayquote}

We now turn from reactivity to strategies for proactivity.
Some communication problems can be anticipated in advance, and documentation is a prime example.

Data scientists are experts at statistics and AI modeling but are typically novices to the business domain, whereas stakeholders are the opposite.
To bridge the gap, one of the practices we observed was when data scientists proactively documented model information to inform their collaborators.
Data1 succinctly summarized the intended audiences for these documents and the goal, \textit{``So I guess (the documentation) is high level business and high level technical. So hopefully regardless of who the person is they can get some kind of a sense of what's happening here.''}

Incorporating domain knowledge into a model is a key to successful modeling. To do so, data scientists ought to understand domain knowledge embedded in their data. Our participants maintained documentation per project to capture domain knowledge.
Unlike AI-concept one-directional documentations for stakeholders, these domain documentations take a role as 
bi-directional channel to be revisited and reviewed with stakeholders. Through this documentation, stakeholders can provide feedback and data scientists can confirm their observations.

Education combined aspects of reactive and proactive strategies.
Although participants were aware of the communication gaps about education, and were thoughtful about how to respond to them, there were still instances where those efforts failed. 
And in the iterations that followed, what started as a proactive activity effectively became a reactive one as participants tailored their content based on the feedback that they got.
It was in learning what stakeholders knew that led to well-designed examples that could be reused effectively.

\subsubsection{\minorrev{Effective Knowledge Sharing}}

\begin{figure}
      \centering
      \includegraphics[width=\linewidth]{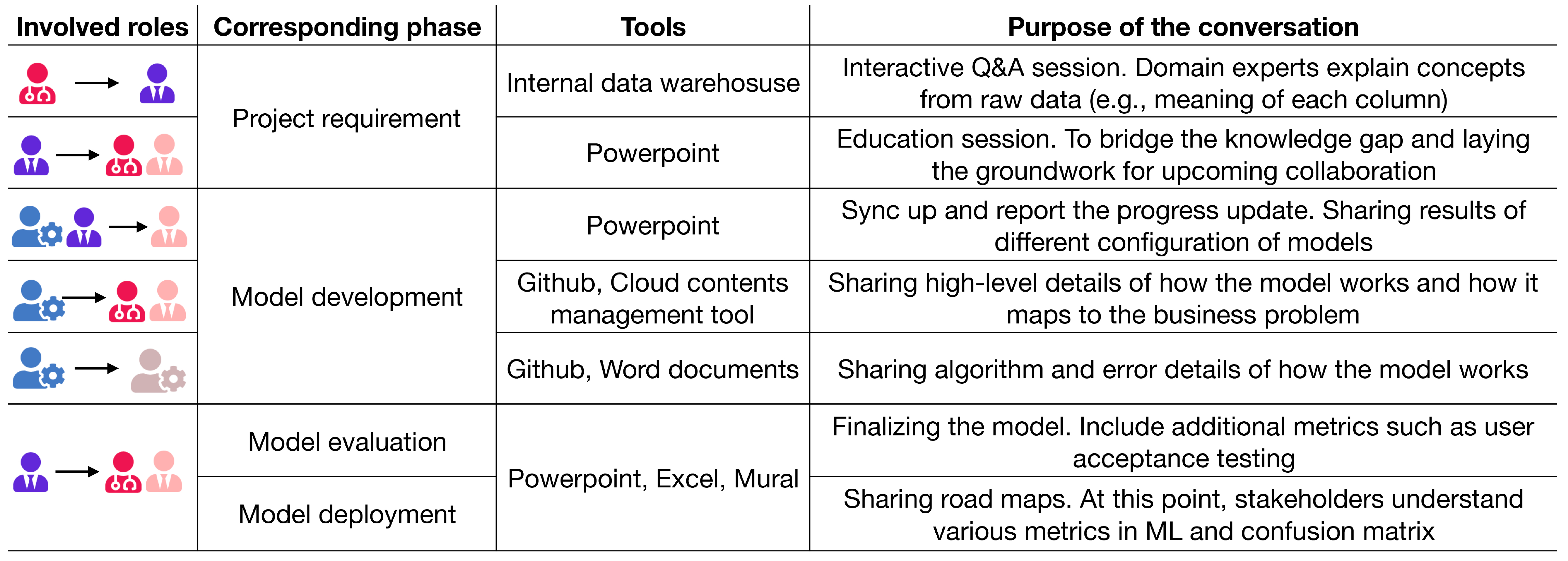}
      \caption{Different tools used for inter-team communication and the reason of conversation}
      \Description{A woman and a girl in white dresses sit in an open car.}
      \label{fig:role-tool}
      \vspace{-5mm}
    \end{figure}

There are lots of different types of information requests that occur, challenges that they carry and approaches to addressing those needs as effectively as possible. Making sure that the information is shared effectively as it is updated is particularly important as much of the work is done independently of others. As we have covered the specific details in the sections above, here instead we provide a summary of our findings. Figure \ref{fig:role-tool} summarizes the inter-role communications that our participants discussed in the interviews. \revision{In the project requirement phase, domain experts, business experts, and strategy consultants gathered together to bridge the knowledge gap and lay the groundwork for upcoming collaboration. Data scientists are usually involved in the model development phase and report the progress on how the model works to domain experts and business experts. They also report low-level technical issues to software developers. Finally, the strategy consultants would report back to the domain experts and business experts to evaluate and deploy the models. In particular, they prefer to use Powerpoint, Excel, and Mural to visually represent information.}

\subsection{Artifact Analysis: Real-World Examples}

\begin{figure}
      \centering
      \includegraphics[width=.98\linewidth]{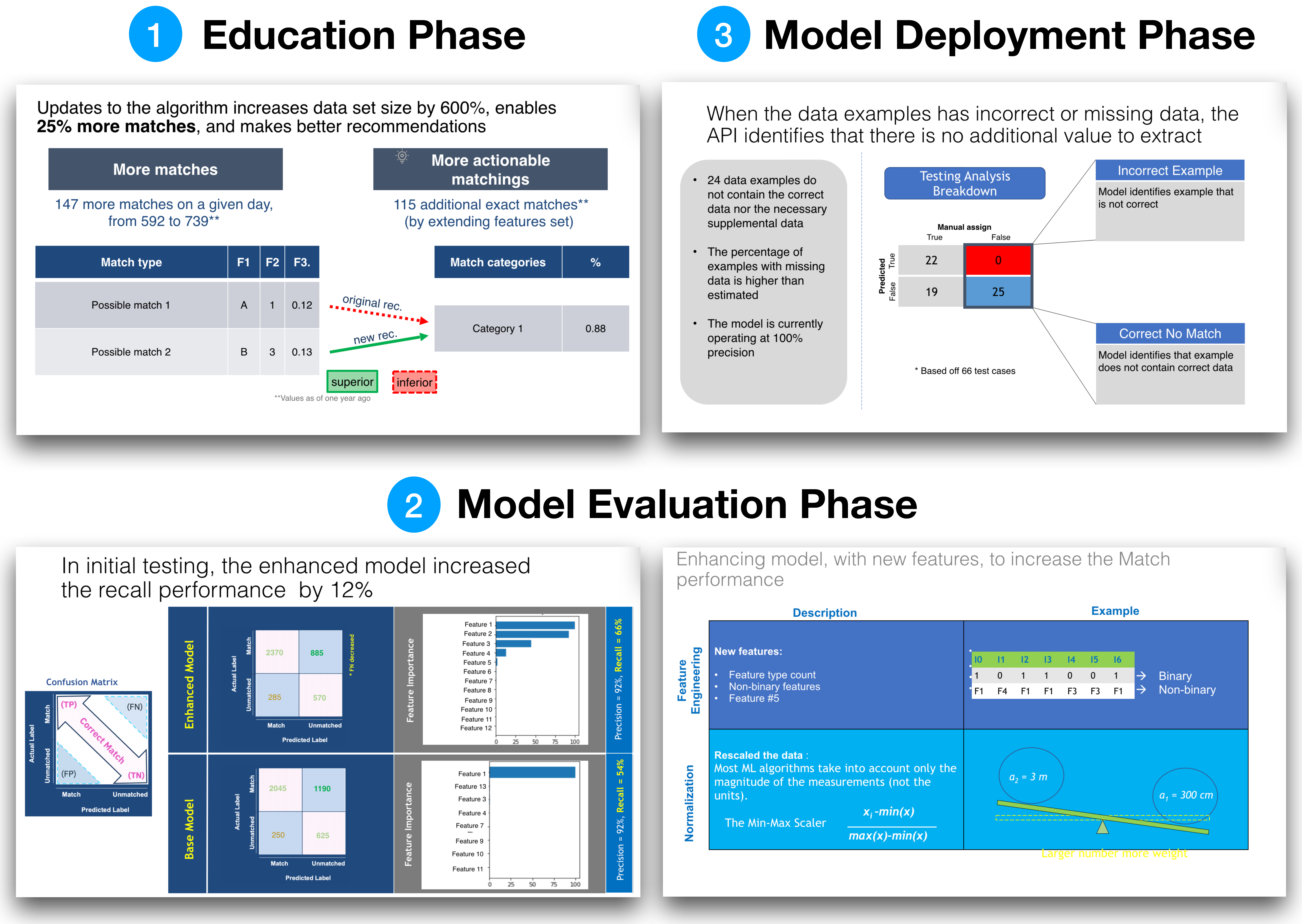}
    \caption{
    Examples of presentation in AI development workflow:(1) Education Phase: The AI team explains basic level of AI knowledge and potential benefits that integrating AI will bring to stakeholders; (2) Model Evaluation Phase: The AI team reports their progress once in a while to stakeholders and reminds them information of AI metrics and presents their model's performance;  (3) Model Deployment Phase: After a few months of collaboration and education, the stakeholders can now interpret AI performance easily. The AI team also shares plans moving forward (Figure has been modified by authors to remove identifiable and confidential information). }
      \label{fig:presentations}
\end{figure}

To triangulate what participants described in the interviews, we also performed an analysis on the artifacts that they shared with us. \revision{In this analysis, we reused} the same codes from Table \ref{tbl:code-list} under the Information Conveyance Selection and \revision{identified} where those approaches were used in the artifacts. Additionally, we wanted to provide real-life examples to the research community of the communication approaches taken to overcome communication gaps across roles.

Figure \ref{fig:presentations} shows a few example slides drawn from slide deck artifacts that participants shared with us.
These decks were created by AI developers to educate stakeholders about AI concepts and to provide status updates on model development progress.
We selected these slides to share because they contain examples of how AI developers convey information in many different ways.
Note that we have modified these slides from their original versions due to confidentiality concerns. 
Slides are referenced according to the AI development stage \revision{in which} they occur and which slide from the set they are in. 
For example, 
in Figure \ref{fig:presentations},
slide \encircle{2}-2 describes the second slide from the Model Evaluation Phase. \minorrev{The collection of the slides is not comprehensive and does not include all the stages with involving different roles across multiple teams (i.e., data requirements, handoff). We focused on only the three stages because those are the stages that data scientists were engaged to share \emph{their} artifact instead of collecting requirements or dealing with logistics.}

In the interviews, participants described the effort they took in creating specialized visualizations to express various kinds of information. 
One way was to explain data science concepts using multiple strategies.
Slides \encircle{2}-1 and \encircle{3} show how a confusion matrix can be used to explain the concepts of recall or true/false negatives.
Additionally, in the upper right of slide \encircle{2}-2, a simple table visualization provides a quick explanation of how features in a model changed from binary to non-binary. 
Feature changes were also visualized in terms of how they influenced the quality of the model.
For example, the right half of \encircle{2}-1 also compares how a different models features changed, and also showcases how much impact each feature has in the final prediction.
Finally, visualization can aid comparison as in \encircle{1} which presents two examples from two different datasets and explains how they both ultimately generate the same prediction.

In some cases, a simple picture conveyed the information for the AI developer.
In slide \encircle{2}-1, on the bottom-right, the analogy of a scale provides an abstraction for the role that normalization plays in this particular model.
Images can likewise be an effective recall mechanism and help trigger memories. 
The small descriptive variant of the confusion matrix on the left side of slide \encircle{2}-1 was used in such a way.
Strat1 explained to us that he would intentionally repeat it in his slides so that people recalled prior discussions about it.
These creative uses of images highlight ways the 
AI team\revision{'s strategies} in overcoming communication gaps.

Outside the specific visual aspects of the slide, we noticed that each slide was highly targeted.
Slides were designed to only deliver a single message that was typically contextualized in the problem domain.
This stands out in headings of the slides and the other explanatory text that they contain as in slide \encircle{3}.

The variety of approaches on display over these four slides helps to demonstrate how precisely targeted and carefully crafted information is by AI developers, before it is used as a tool to cross a communication gap. The team must mold its approach to best suit its particular audience. 

Our findings for the artifact analysis align with what participants talked about in the interviews. 
There is no one-size-fits-all approach here. 
Designing information content was time-consuming, yet necessary work as it is critical to building an effective SMM and to the project's success. 
Not all content needed to be bespoke, however.
As is the case in slide \encircle{2}-1 with the reminder version of the confusion matrix, we also identified some evidence of reuse.
For example, the confusion matrices themselves were used across multiple presentations, even if they were used to answer slightly different questions. \revision{Participants also stated that they would reuse explanatory content from past artifacts, adapting the content to fit their current context. Yet, such reuse relied on participants recalling ideas from the past, as none of the tools provided affordances to search for these kinds of explanations.}
Given the small scope of this case study, we cannot generalize beyond \revision{these participants}, but at least for these observations, AI developers tended to iteratively customize a piece of information until the communication gap is crossed, and then \revision{if they remembered}, reuse it if possible in the future.

\section{Discussion}

Our research found that AI developers face various communication challenges and devise their own specific responses to these challenges. We discuss what are the best current practices that we observed during our study, where tools can help improve the collaboration experience, and how machine teaching and automated ML can lead to efficient communication among teams. We also describe the limitations and future work of this study.

\subsection{Best Practices in AI Development Workflow}

\begin{table}[t!]
\caption{A list of codes used for best practices (Top) and identified practices and corresponding SMM practices and components (Bottom) } 
    \begin{tabular}{p{4cm}p{9cm}}
    \toprule
    \textbf{SMM practice}  & \textbf{Definition} \\ 
    \midrule
    Planning         &  In the beginning of the project, team members spend time to brainstorm and define a goal of the project \\ 
    \midrule
    Team-interaction training          & Team members share embedded information in team tasks \\ 
    \midrule
    Reflexivity         & \emph{Team members} work together to reflect upon previous experience and mistakes                                 \\
    \midrule
    Self-correction training        & \emph{Individual team member} works to reflect upon previous experience and mistakes        \\ 
    \bottomrule
    \end{tabular}
    \label{tbl:code-list}
     \vspace*{5mm}
    
     \begin{tabular}{p{4cm}p{4.5cm}p{4.5cm}}
    \toprule
    \textbf{Practice}  & \textbf{SMM practice} & \textbf{SMM components} \\ 
    \midrule
    Spending time to understand the problem and iterate         &  planning, \newline self-correction training &
    task procedure, task strategies, environmental constraints \\ 
    \midrule
    Capturing frequently asked questions by stakeholders & team-interaction training & information sources \\ 
    \midrule
    Documenting domain knowledge in data          & team-interaction training   & information flow, \newline communication channel                              \\
    \midrule
    Running small pilot studies to evaluation models in development       & self-correction training, \newline team-interaction training  &
    task procedure, task strategies\\ 
    \midrule
    Rolling from one business sector to another       & reflexivity  &
    task procedure, task strategies\\ 
    \bottomrule
    \end{tabular}
\end{table}

In this section, we consolidate observations regarding the best practices that AI developers observed during our study. 
To understand best practices, we explain emergent practices through a lens of SMM. 
To better understand why the best practices were successful, we borrow from Yu and Petter's work on how Agile development practices map to successful mental models~\cite{yu2014understanding} which are defined in the top half of Figure \ref{tbl:code-list}. 
We then frame AI developers' best practices from this perspective to provide additional explanation for why these practices are successful as shown \minorrev{in} the bottom half of Figure \ref{tbl:code-list}.

\subsubsection{Longer Planning Period than Actual Implementation}
Software development teams tend to spend extensive time on planning and brainstorming~\cite{myers1992survey}.
Although AI application development is known to be unique and different from software engineering~\cite{amershi2019software}, we discovered a commonality between two domains \minorrev{at a high-level}. Both ML and software engineering spend a great deal of time on planning and brainstorming. Through this period, AI teams are able to learn about domain knowledge and adjust their implementation accordingly (self-correction):

\begin{displayquote}
\textit{``We spent around four months just doing the query work, understanding those business scenarios, and then we held a couple interviews with sales rep to confirm our observations. Based on the knowledge, ... we started to do initial models ... it's more like an incubation. Then, we held what we called garage session---a three-day workshop with assets team''}
[Data3]\end{displayquote}

\subsubsection{Shared \minorrev{Documentation} Effort to Build External Knowledge}

As we stated earlier, managing shared information storage for team members is key to successful communication. It was shown that (1) it is helpful to answer repeated questions from stakeholders (2) it becomes an effective communication channel to bridge between teams. 

It is worth noting how well documentation is utilized across different teams in our AI development teams --- unlike previous literature claimed.
The research literature revealed that data scientists do not use documentation to record their data-management or project plans~\cite{pine2015politics}, and that they do not use documentation systematically throughout a data science lifecycle~\cite{zhang2020data}, and that they do not use documentation resources enough~\cite{rule2018exploration}. 

We believe this difference occurs due to the team structures of our participants: two independent teams that have not previously collaborated, but they form a team in an ad-hoc manner to integrate ML and AI in their business model. Therefore, the two teams have not fully understood each others' expertise or established collaboration protocols. By using documentation as a conversation tool, teams can have contextualized and informative communication, and at the same time conduct knowledge management (team-interaction training).

\subsubsection{Rolling from Small Pilot Studies to Bigger Projects}

We observed that AI teams follow agile practices in software engineering. Data scientists try to minimize time to deliver an ``executable'' version of the model to stakeholders. The rationale appear to be that this helps AI teams to keep stakeholders in-the-loop and gain insights to tweak model behavior (self-correction). Additionally, it is also an opportunity to showcase their progress and the ability of the model:
\textit{``The goal is to understand their experience and make sure they [stakeholders] feel like our optimum (value) is correct and feel they are confident using the model''} [Data2].

Once an AI team successfully launches a model, they slowly expand the model from one sector of the business to others. As the model is stable in one sector, they can streamline the expanding process based on previous experiences in modeling (reflexivity).
 
\subsection{Design Implications \minorrev{for Collaboration Tools}}
Our case study reveals unique challenges in the ability of AI teams/devs to communicate across project roles. Consequently, we propose the following improvements to current collaborative tools.

\subsubsection{Customizing Documentation to Enable Cross-Role Communication}
Information and how it is presented needs to be highly tailored for the roles receiving that information or, simply put, one size does not fit all. Participants spent a lot of their time creating visualizations, selecting examples and creating stories to share information effectively.
Documentation tools that  capture and share information to different roles along the AI lifecycle likewise need to tailor their content and presentation to the needs of the role.
Existing efforts around documenting aspects of the AI lifecycle in dataset documentation~\cite{gebru2018datasheets} and model service documentation~\cite{mitchell2019model, arnold2019factsheets} are fairly static and do not consider questions of how to dynamically vary the content that is displayed for the role in question.
One exception is Hind et al.'s formative work on difficulties users have creating such documentation, which also advocated for user-centric output~\cite{hind2020experiences}.
Providing a tables of statistics or visualizations without context is not going to address the needs of these users.
To facilitate cross-role communication on these teams, \minorrev{documentation tools} need to be aware of how these roles process information, what information they care about and the best ways to share this information with them.

\revision{Currently, data scientists are left adapting imperfect tools to their needs. For example, each participant used PowerPoint for communication or educational needs. Yet participants also made clear that PowerPoint is not well-suited to creating and maintaining complete and timely documentation. Instead, participants described using it to capture a snapshot of the current state of the work. Recall that creating that snapshot can be a non-trivial amount of work. Strat1 said it took ``2--3 days'' to create slides. 

So why do they use PowerPoint given its high cost? Although we did not ask explicitly, the interviews with participants suggest that the flexibility and familiarity of PowerPoint may be part of the reasons why. This may be particularly important for the transfer of knowledge from AI Developers to Stakeholders - especially if the Stakeholders are planning to re-use content from the AI Developers' presentation. We remember that Data1 reported that, 
\textit{``...the major outcome [was] just try to get management up to speed... and hopefully also get them excited about it...''}
If materials are going to management, who may send those materials on to other management, then the use of a generic presentation tool may be strategic.

We speculate that despite the sometimes high cost of translating content to slides, the benefit of using slides is much higher still. This high-cost situation suggests a need for closer integration between data science tools and communication tools, echoing prior literature~\cite{zhang2020data}. One example of such integration is Callisto which brings together chat and computational notebooks \cite{wang2020callisto}, but more research is needed in this direction.
}

\subsubsection{Towards Automatically Shared Mental Model Updates}
Our interview study found that AI developers mainly shared the implementation progress through an external medium (e.g., PowerPoint, PDF document) other than their working environment (e.g., code editors, computational notebooks).
Not only is it a time-consuming process for them to decide what to present and how to best deliver the results, it is impossible for them to keep up to date with the rapid changes that occur during the AI development pipeline.
One possible solution to address this issue is through the use of automatic summarization tools built on recent advances on code summarization and explainable artificial intelligence.
Such tools could monitor inter-team changes to code and other information reposoitories to keep team members up to date with the latest changes.
SMMs suggests that enhanced visibility may enable better collaboration since everyone would be more aware of the recent changes; however, there would still be open questions on perceptions of trust, role-specific approaches to filtering and presenting information, and how such tooling can be designed as to not be yet another distraction that teams have to deal with.

\subsubsection{\minorrev{Beyond ``Game of Telephone'': Re-imagine Efficient Communication without Mediators}}
We observed \revision{that} communications among AI teams often involve indirect feedback to AI models. Many forms of conversation involve many different roles.\footnote{\revision{We attached the team composition in the Appendix \ref{appd:team}.}} For example, in the project requirement stage, domain experts explain the domain to strategy consultants, which strategy consultants later \emph{translate} to data scientists. Hou et al.~\cite{hou2017hacking} identified this role and referred to this human intermediary as a \emph{coordination broker}, who bridges between technical and non-technical roles. 

This generates a situation of the \emph{Game of Telephone} -- i.e., additional mediators have to be involved to incorporate their domain knowledge~\cite{pinhanez2019machine, 10.1145/3311957.3359512}. We envision that with the right tool and education, we can take it to a more efficient way of conversation and avoid the game of telephone. Recent work in AutoML and machine teaching makes promising progress~\cite{wang2019hai,Wilder2020complement, cambronero2019autogenerating,autods}. With this technology, non-ML experts can build and provide insight into models without additional mediators.

\revision{

\subsection{Threats to Validity, Limitations, and Future Work}

Given the small scope of this study and its qualitative research nature, the findings from this paper are preliminary. We caution the readers to consider the following limitations, if they plan to generalize the paper's findings outside the investigated context. 

We recruited informants only within a single company \minorrev{(IBM)}, where the first author works. Because our interviews involved private and confidential project details, the informants needed this privacy protection in participating in this study, as we have noted earlier in section \ref{subsect:participants}. The results reported in this paper may be biased by this restriction of informants recruited. 
AI teams at other organizations may be structured differently; have a different composition of roles; or may be part of the Stakeholder team themselves. Such teams may not face some of the communication gaps we described or may face different ones entirely. Future work may explore whether the same types of issues would be seen in other types of organizations outside the team that we studied.

Another limitation is that we had only four interviews and 10 interview sessions. Despite the depth of each of those interviews, our findings should be considered formative, until further studies with a larger and different informant sample reaffirm our findings. Ideally, we would also like to triangulate through methods such as large-scale surveys of teams and/or on-site or online ethnography method. If we collect enough data on both successful and failed communication practices, maybe we can even apply a machine learning algorithm to learn what practices lead to successful communication.

Furthermore, we acknowledge that communication in an AI project is bi-directional, where both AI developers and stakeholders may drive and participate different parts of communication in an AI project~\cite{zhang2020data}. However, the people whom we interviewed represent only one perspective on the projects that they discussed. Thus, we remind the reader that we presented a one-sided perspective on the problems represented. Future work should focus on the complementary perspectives from the Stakeholder teams.

Last but not least, we are also aware that a theoretical lens both magnifies items in its focus, and marginalizes other items. Our choice of SMMs was useful to clarify certain issues, but we anticipate that a future study may be able to use multiple theory-lens for a broader view of communication issues. As an example of the selected model's limitation, our interviews did not explicitly \minorrev{tag along} the SMM's coordination principal. This could have been an effect of our interview design or our focus on the model evaluation and deployment phases. We speculate that much of the coordination activities are established in earlier phases such as project requirements and data requirements and gathering as these phases represent the earliest interactions between the AI team and the Stakeholder team~\cite{zhang2020data}. Had we focused our study on these earlier phases, it is reasonable to expect that more coordination-principle-related details would have emerged. 
In the future, we also plan to use common ground theory \cite{clark1991grounding} and theories of small group activity, as recently reviewed by Lee and Paine \cite{lee2015matrix} and by Stahl \cite{stahl2013theories}.

Despite these listed limitations, we believe that our results and our methodology in this paper will be useful for future work. As Paul Dourish argued ~\cite{dourish2014reading}, a qualitative work's validity and power lies together with the other qualitative works (e.g.,~\cite{hou2017hacking, mao2019data,muller2019data}). These various accounts from different contexts and different perspectives can form a systematic understanding of the communication practices in AI teams. 

}

\section{Conclusion}
In this paper, we presented a case study on how AI developers overcome communication challenges across roles in the AI development lifecycle.
With the aid of a shared mental model lens, we identified the key communication gaps that AI developers faced.
These gaps had an overarching theme of educating others and specifically were about gaps in knowledge, establishing trust and setting expectations.
We also shed light on how and why AI developers cross communication gaps.
We extracted rich details from the interviews that highlighted the many different information requests that AI developers face and also shed light on how they approach and fulfil them.
Together these results informed some best practices in AI development workflow and provide important insights to researchers and designers who are interested alleviating some of the communication challenges that AI developers face.
Until then, all participants can do is wait for a magic tool to help them overcome their communication gaps.
\begin{displayquote}
\textit{``I've been working with this project for so long, I could easily kind of dump [information about] it out. It's just a matter of how do you go about this? ... if there was some kind of template or tool to show me exactly what I need to make, ... that would have saved me a lot of brainstorming.''} [Strat1]
\end{displayquote}

\bibliographystyle{ACM-Reference-Format}
\bibliography{main}

\newpage

\appendix
\section{Team Composition of Each Team}\label{appd:team}

This shows our interviewees' team configurations and how different roles are involved in each stage of \ac{ML} workflow. The team's composition varies due to special occasions, such as data availability, data already being labeled, and different role having responsibility across different teams.  
An aggregated view can be found in Figure~\ref{fig:workflow-role}.

\begin{figure}[h]
      \centering
      \includegraphics[width=.55\linewidth]{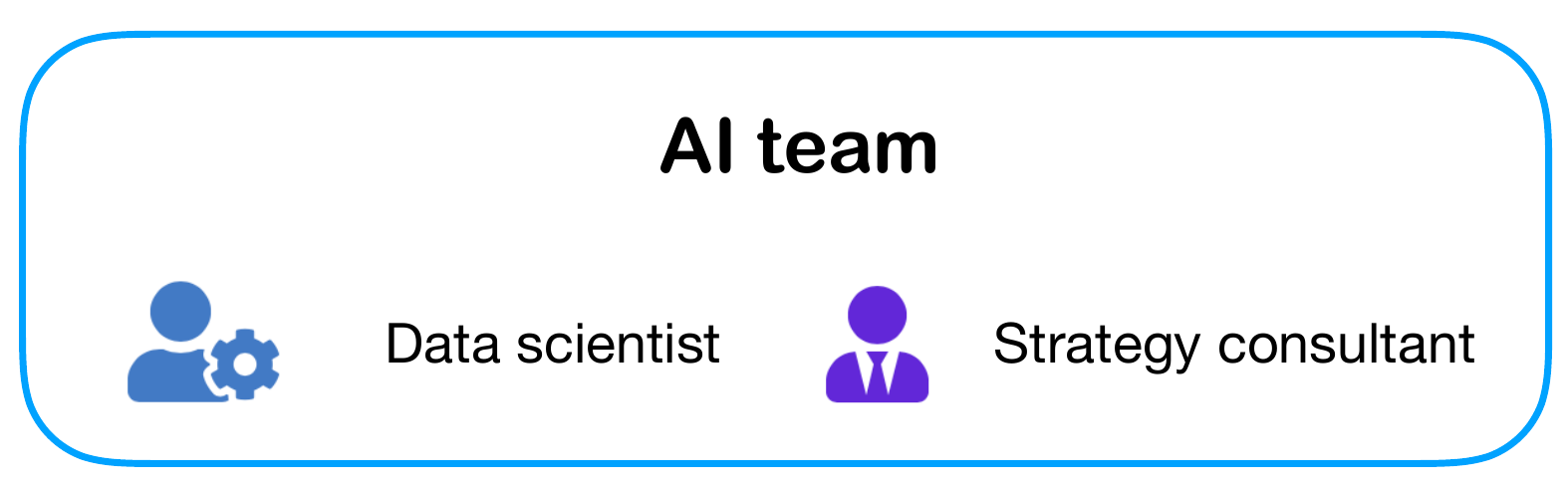}
      \label{fig:data1-team}
    \end{figure}
    \vspace{-8mm}
    \begin{figure}[h]
      \centering
      \includegraphics[width=\linewidth]{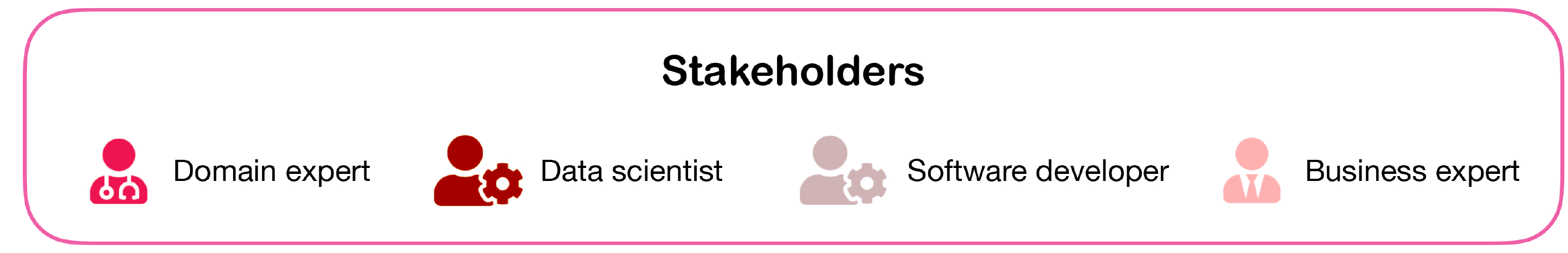}
      \label{fig:data1-team}
    \end{figure}
\vspace{-8mm}
\begin{figure}[h]
      \centering
      \includegraphics[width=\linewidth]{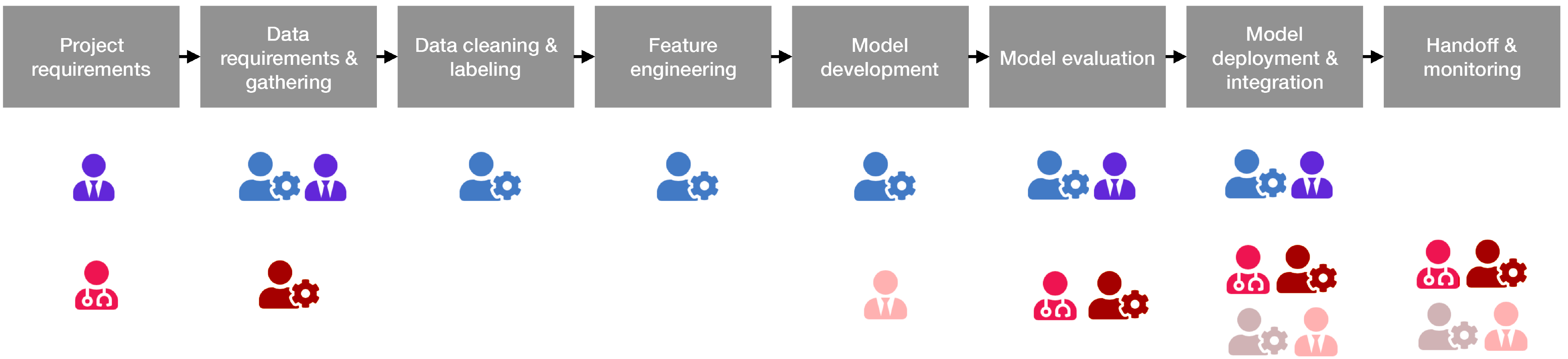}
      \caption{Data1 team}
      \label{fig:data1-team}
    \end{figure}

\begin{figure}[h]
      \centering
      \includegraphics[width=\linewidth]{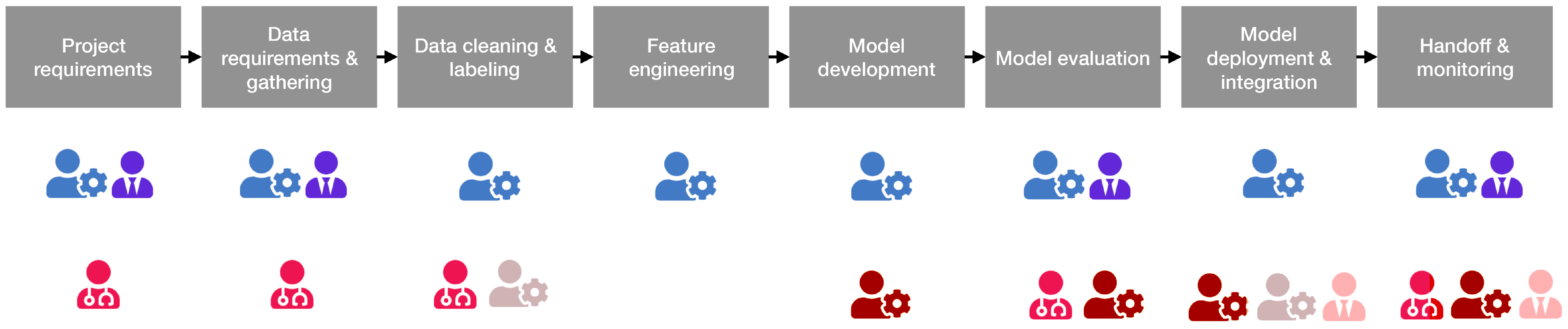}
      \caption{Data2 team}
      \label{fig:data1-team}
    \end{figure}
    
    \begin{figure}[h]
      \centering
      \includegraphics[width=\linewidth]{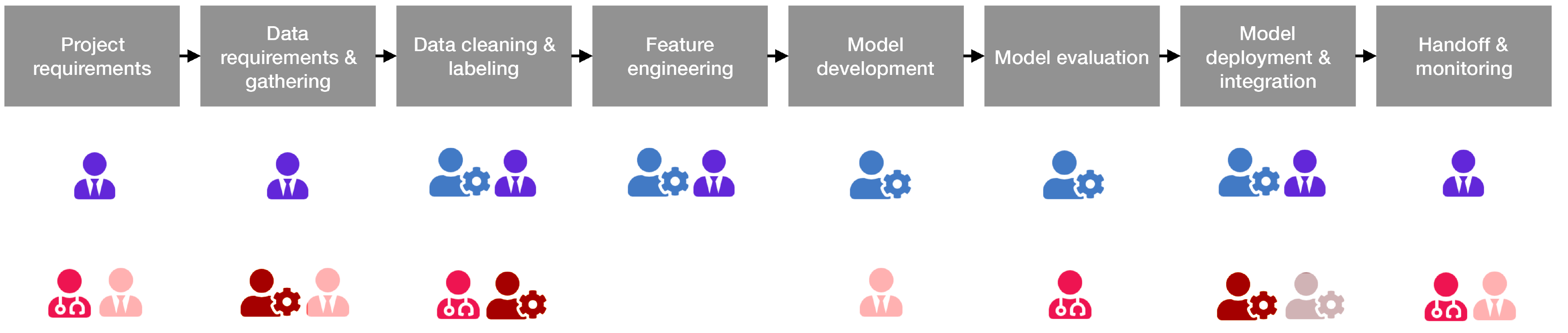}
      \caption{Data3 team}
      \label{fig:data3-team}
    \end{figure}
    
    \begin{figure}[h]
      \centering
      \includegraphics[width=\linewidth]{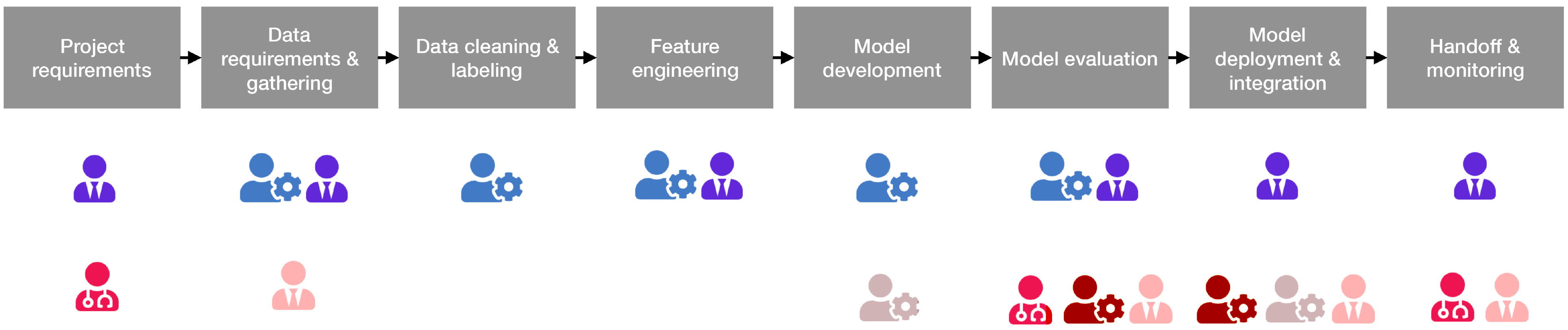}
      \caption{Strat1 team}
      \label{fig:strat1-team}
    \end{figure}

\end{document}